\documentclass[%
 aip,
 jmp,%
 amsmath,amssymb,
 reprint,%
]{revtex4-1}
\usepackage{float}
\usepackage{graphicx}
\usepackage{dcolumn}
\usepackage{bm}

\begin{document}

\preprint{AIP/123-QED}

\title{The effect of randomness for dependency map on the robustness of interdependent lattices}

\author{Jing Yuan}
 \affiliation{State Key Laboratory of Networking and Switching Technology, Beijing University of Posts and Telecommunications, Beijing, 100876, China}

\author{Lixiang Li}%
 \email{lixiang\_li2006@163.com}
 \affiliation{State Key Laboratory of Networking and Switching Technology, Beijing University of Posts and Telecommunications, Beijing, 100876, China}
\author{Haipeng Peng}
 \affiliation{State Key Laboratory of Networking and Switching Technology, Beijing University of Posts and Telecommunications, Beijing, 100876, China}
\author{J\"{u}rgen Kurths}
\affiliation{Potsdam Institute for Climate Impact Research, Potsdam, D-14473, Germany}%

\author{Xiaojing Hua}
\affiliation{School of Science, Beijing University of Posts and Telecommunications, Beijing, 100876, China}%

\author{Yixian Yang}
 \affiliation{State Key Laboratory of Networking and Switching Technology, Beijing University of Posts and Telecommunications, Beijing, 100876, China}
\date{\today}

\begin{abstract}
For interdependent networks with identity dependency map, percolation is exactly the same with that
on a single network and follows a second-order phase transition, while for random dependency,
percolation follows a first-order phase transition. In real networks, the dependency relations between
networks are neither identical nor completely random. Thus in this paper, we study the influence of
randomness for dependency maps on the robustness of interdependent lattice networks. We introduce
approximate entropy($ApEn$) as the measure of randomness of the dependency maps. We find that there is
critical $ApEn_c$ below which the percolation is continuous, but for larger $ApEn$, it is a first-order
transition. With the increment of $ApEn$, the $p_c$ increases until $ApEn$ reaching ${ApEn}_c'$ and then
remains almost constant. The time scale of the system shows rich properties as $ApEn$ increases.
Our results uncover that randomness is one of the important factors that lead to cascading
failures of spatially interdependent networks.
\end{abstract}

\pacs{89.75.Fb, 89.75.Hc}
\keywords{interdependent lattices, randomness, dependency map}
\maketitle

\begin{quotation}
The interdependent networks which fully consider the interactions between networks have been used to
model real complex systems better. Robustness is one of most important properties for interdependent
networks especially spatially interdependent networks, since most of the infrastructure networks are
spatially networks. In real interdependent networks, the dependency
relationship are not usually random. Thus we analyze how the randomness of dependency map
affects the robustness of interdependent lattices which are used to model the spatially interdependent networks.
We find that the randomness of dependency map between networks is quite critical for the robustness of interdependent lattices.
\end{quotation}

\section{Introduction}

Robustness is one of the most important properties of complex networks and has been
widely explored on single networks in the last decade\cite{newman2003structure,callaway2000network,cohen2000resilience,cohen2001breakdown,cohen2002percolation,albert2000error,vazquez2003resilience,crucitti2003efficiency}.
However, complex systems are rarely isolated. The more casual situation is that networks
usually interact with other networks such as transportation networks and financial systems\cite{gao2012networks,cardillo2012modeling,criado2007efficiency,criado2010hyperstructures,cardillo2013emergence}.
In the case of interdependent networks, conclusions are often far different from single networks.
In particular, a removal of a very  small fraction of nodes can lead to catastrophic failures on the whole network\cite{buldyrev2010catastrophic}.
A theoretical framework based on percolation theory has been established to analyze the resilience of interdependent systems\cite{gao2012networks,danziger2014introduction},
and much details have been explored\cite{boccaletti2014structure,berezin2015localized,buldyrev2011interdependent,parshani2010inter,cellai2013percolation}.
The fraction of interdependent nodes is one important factor that will influence the phase transition of the networks\cite{kornbluth2014cascading,parshani2011critical}.
Also, the overlap of links can significantly change the properties of the percolation, and there is a critical point above
which the emergence of the mutually connected component is continuous\cite{cellai2013percolation}. The
presence of degree correlations in multiplex networks can modify drastically the
percolation threshold\cite{buldyrev2011interdependent,parshani2010inter}.

    Most previous models have focused on interdependent random and scale-free networks in which
space restrictions are not considered. However, many real-world systems such as power grid networks
and computer networks are embedded in two-dimensional space\cite{bashan2013extreme,danziger2013interdependent}.
In interdependent random and scale-free networks, the overlap of links and degree correlations
will change the properties of phase transition. Nevertheless for spatially embedded interdependent networks
which are modeled as square lattices, the overlap of links or the degree correlations of nodes lose their
significance, since their network topologies are identical. The spatially interdependent networks are
extremely vulnerable. Any fraction of interdependent nodes will lead to first order transition\cite{bashan2013extreme}.
From an identical dependency map to totally random dependency map, the randomness of the dependency map may be one of the
most important factors leading to the emergence of discontinuous percolation.
In most real interdependent systems, dependencies are neither totally random nor identical.
Research on the resilience of intermediate systems that lie somewhere in between these two extremes
is of high practical significance and needs further exploration.

	From this perspective, we study the relationship between the dependency's randomness and stability of
the system of two interdependent spatially embedded networks.
We use approximate entropy($ApEn$) as the measure of randomness. One of the big challenges here is how to introduce
controlled degree of randomness into the system. Therefore, we propose an intermediate model which describes
the system with dependency map between identical map and totally random map.
Inspired by the constructing procedure of the Watt-Strogatz small-world model\cite{watts1998collective}, starting from an identical dependency map,
we rewire each dependency link at random with probability $q$.
By increasing $q$ from $0$ to $1$, the $ApEn$ increases monotonically. Therefore, the traverse of randomness
can be generally represented by $q$. We reveal that there is a critical value $q_c$,
for which the percolation transition becomes continuous, whereas for any $q > q_c$ the collapse is discontinuous.
Changing the topologies on a single layer, we discover that  $q_c$ is different for interdependent
scale-free networks, Watts-Strogatz networks, and Erd\H{o}s-R\'{e}nyi networks.
The critical threshold increases with $q$ when $q<q_c'$ and remains approximately constant when $q>q_c'$.
Additionally, we present an analytical method for time scale of cascade failures based on critical p and find that
the four topologies display rich transient properties when $q$ changes from $0$ to $1$.
Finally, we analyze the influence of limited dependency length on spatial networks. We show that
with the same dependency length, a linearly dependent system is always continuous, but not continuous for
some locally randomly dependent system. Our results show that the randomness of  dependency may be one of important factors for
extreme vulnerability of spatially interdependent systems.

\section{Model Description}

Our model of interdependent networks is realized via  two networks($N = 10^6$)
A and B under full dependency. Here one network is the
copy of the opposite network and their average degree $<k> = 4$(the same as a square lattice).
The degree distribution of the scale-free network is $<k>^{-\lambda}$ where $\lambda = 3.0$.
In each network, each node has two types of links: connectivity
link and dependency link. Also every node in network A is connected with one and only one  node in network B.
For a square lattice, each node is connected to its four
nearest neighbors within the same lattice via connectivity links.
All dependencies in our model are mutual and bidirectional.
Dependency is taken to mean that if a node in network A is removed from
the system and a node in B that depends on it will be removed from B as well.
Thus failures of nodes iterate until mutually connected giant component of both networks emerges.
This process is called cascade failures and see Methods for details of cascade process of the system.

\begin{figure}
    \centering
    \includegraphics[width= 8.4cm]{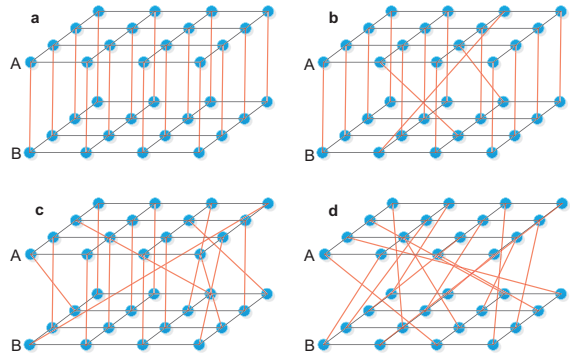}
    \caption{The interdependent square lattice with the rewiring probability of dependency links
     $q = $  0 , 0.25, 0.50, 1.00 respectively. When $q =0$,
    the dependency map is an identical mapping i.e.  node $i$ in network A is dependent on
    node $j$ in network B where $i = j$(Fig.1.\textbf{a}) .  When
    $q =1.00$, the situation is the same as totally random mapping(Fig.1.\textbf{b}). When $q = 0.25~or~0.5$
    , the situation is between both extremes(Fig.1.\textbf{c,d}).}
\label{fig1}
\end{figure}

There are two extreme situations. i) node $i$ in A  depends
on node $j$ in the B  such that $j = i$. We call it identity dependency map(Fig.\ref{fig1}.\textbf{a}).
ii) The random dependency map as most papers considered(Fig.1.\textbf{b}).
Like the constructing procedure of the Watt-Strogatz small-world model, starting from
the identity dependency map, we rewire each dependency link at random with probability $q$,
while guaranteeing that each node in A depends on one and only one node in
B($0\leq q \leq 1$). We sample $q = 0,0.25,0.50,1$ and plot them in Fig.1.

\begin{figure*}
    \centering
    \includegraphics[width = 17cm]{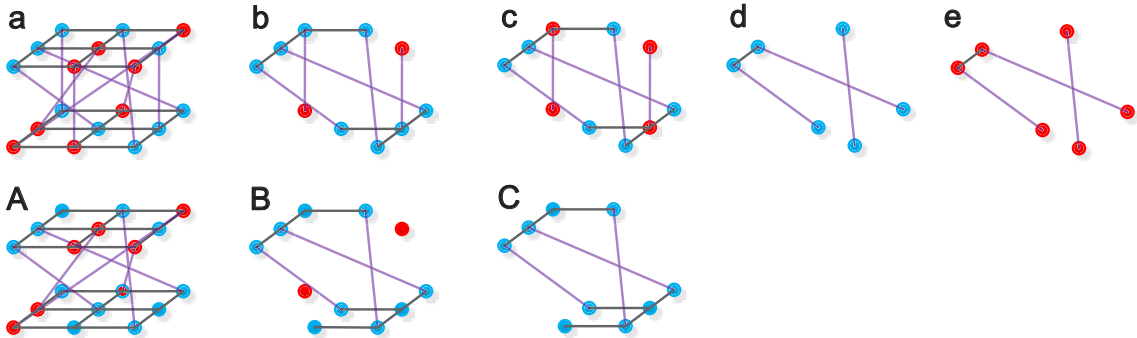}
    \caption{Difference of cascade failures between partially randomly interdependent lattices and partially
    interdependent lattices. The blue points stand for the survived nodes, while the red points stand for the
    failure nodes. Figure sequence a-f stands for the cascade failures process in partially randomly
    interdependent lattices with $q = 5/9$(fraction of nodes that are randomly dependent and the remaining
    $1-q$ of nodes are dependent with the identical nodes in the opposite network) and $p = 4/9$(fraction
    of nodes initially removed). Figure sequence A-C stands for the cascade failures process in partially
    interdependent networks with $q = 5/9$(fraction of nodes that are dependent and the remaining $1-q$
    are autonomous)and $p = 4/9$. It can be obviously seen with the same $q$, the size of giant component
    in partially randomly interdependent lattices is $0/9$, while the size of giant component in partially
    interdependent lattices is $4/9$. The cascade failures process differs for these two models.}
\label{fig2}
\end{figure*}

Note: We must figure out that our model is different from partially interdependent lattices proposed by
Amir Bashan et al\cite{bashan2013extreme}. In partially interdependent lattices, there are interdependent
lattices with a fraction $q$ of interdependent nodes and
the remaining $1-q$ of nodes autonomous. In our model, however, the remaining $1-q$ nodes are
connected with the identical nodes in opposite network.
The subtle difference between these two models will lead to huge difference in cascade failure process.
We will illustrate it in Fig.\ref{fig2}. In Fig.\ref{fig2}, we can see that
the cascade failures process differs much between these two models: with the same $q = 5/9$ and $p= 4/9$, the size
of the giant component in our model is $0/9$, while the size of giant component in partially interdependent networks
is $4/9$. This apparently show that the our model is different partially interdependent lattices.

\section{Results}
Entropy can be used to measure the randomness effectively\cite{pincus1994physiological}.
In fact, approximate entropy($ApEn$) is adopted in this paper for computation convenience.
When $q = 0$, $ApEn$ is nearly 0, and when $q = 1$, it reaches its maximum.
The $ApEn$ is a continuous and monotonically increasing function of $q$(Fig.3).
However the randomness is not fully represented by rewiring dependency links,
since the locally randomly interdependent lattice\cite{li2012cascading} in which $q = 1$ is not
totally randomly interdependent but with length constraints.
Then considering a more casual situation, the permutation of $1 \sim N$ cannot be exhausted by
rewiring the dependency links of identical mapping at probability $q$.
But as the approximate entropy changes continuously with $q$,
we can change $q$ to generally represent all approximate entropies.
\begin{figure}
\includegraphics[width= 8.4cm]{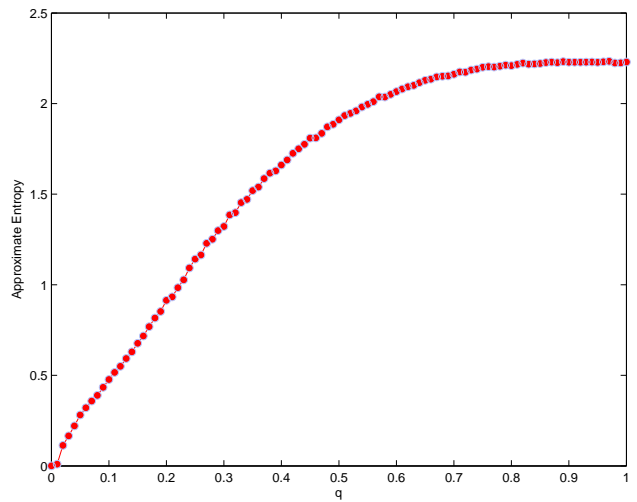}
\caption{ The value of $ApEn$ under different $q$.
 When q = 0, $ApEn$ is nearly 0,and when q = 1, $ApEn$ reaches its maximum. The $ApEn$ is a continuous function
of q and it change monotonously as the increment of q.}
\label{fig3}
\end{figure}

Through simulation, we find that there is a critical $q_c \approx 0.13$ for an interdependent lattice network
below which the percolation is second-order but first-order above.
In Fig.4, we can see that for $q=0.1$, the decrease of giant component occurs in multiple steps,
characteristic of a second-order transition. For $q = 0.2$ and $q = 1.0$, the giant component may completely collapse by
removal of  a small fraction of nodes, characteristic of a first-order transition(Fig.4).
$p_c$ increases with $q$. When $q$ is relatively small, $p_c$
increases approximately linearly. And when $q > q_c'$, $p_c$ remains almost constant,
i.e. as more random the dependency is, as more fragile is the system.
Analogously, for interdependent scale-free networks, Watts-Strogatz networks and Erd\H{o}s-R\'{e}nyi networks,
there is also a critical $q_c$.
We find that $q_c^{sl} < q_c^{SW} < q_c^{ER} < q_c^{SF}$.
Additionally  $p_c$ of lattice networks is generally greater than that of other networks.
Smaller fraction $1-p_c$ will lead to cascade failures. This means that an interdependent
scale-free network is most robust under random attacks, while the square lattice is
most vulnerable. A random network is more stable than a small-world one.

The time scale of cascade failures, i.e. the time that the interdependent networks
need to collapse to the stationary state is one evidently important merit for system's resilience.
The number of iterations( NOI) increases and reaches its peak at the critical $q_c$ and goes quickly
down to a small value with $p$ (Fig.5). So the NOI at $p_c$ is an effective measure for
time scale of the system. The NOI at $p_c$ (i.e.$p_c^q$) is a function of $q$. $p_c^q$ increases
quickly with $q$ when $q < q_c$ and declines very gently above $q_c$ for interdependent
lattice networks. For a scale-free network, it increases until $q_c$ and then starts to decline.
For random networks and small-world networks, it increases monotonously with $q$ but the variation
tendency becomes gentle above $q_c$(Fig.6). All four interdependent systems have variation tendency's changes
around $q_c$. $p_c^q$ of interdependent lattice networks is greater than those of scale-free, small-world  and random networks when $q< q_c^{sl}$. $p_c^q$ of
interdependent square lattice is smaller than those of all other three network types when $q > q_c ^{sf}$.
The NOI reflects the time scale of system collapse. Our results show that the transient characteristics
of the four systems go through rich changes with the variation of $q$.

\begin{figure}
\includegraphics[width= 8.4cm]{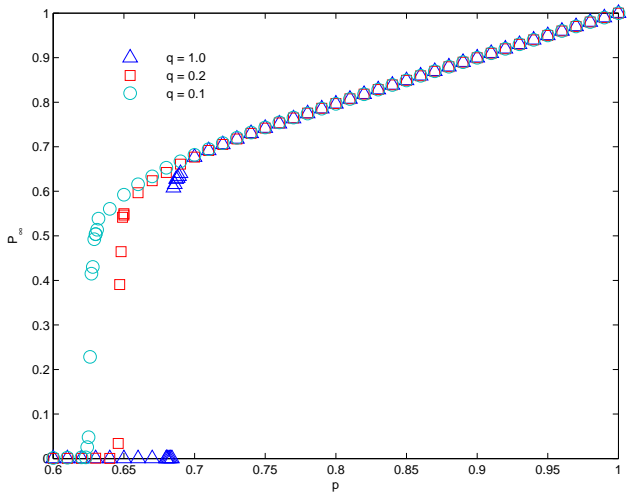}
\caption{Relations of the size of $p_{\infty}$ at steady state
    after random failure of a fraction $1-p$ of the nodes on interdependent square lattices.
The green circles, red squares and  blue triangles for $q = 0.1$,$q = 0.2$ and $q = 1.0$ respectively.
    The numerical results are obtained by averaging 100 realizations of networks consisting of $N = 10^6$.
    For $q = 0.1$, the phase transition of the system is continuous since the giant component occurs
    at multiple steps. For $q = 0.2$ and $q = 1.0$, however, the transition is first-order as
    the giant component collapses even by removing a very small fraction of nodes.}
\label{fig4}
\end{figure}

Finally we check locally interdependent networks in which the distance between two interdependent
nodes is limited($d \leq r$ i.e. $|x_1 -x_2| \leq r $ and $y_1 - y_2| \leq r$ in reference\cite{li2012cascading}).
For simplicity, we consider one more special condition.
Here we split the whole network into $r * r$ small blocks and dependency links are randomized within each block.
We find that there is critical distance $r_c \approx 25$ under which
the percolation is continuous but discontinuous above $r_c$(Fig.7). The $r_c$ is greater than $r_c^{'}$ in \cite{li2012cascading}
because the randomness(approximate entropy) here is lower than that in \cite{li2012cascading}with the same distance.
The corresponding approximate entropy $ApEn \approx 0.923$ of $r_c$ is approximately
equal to $ApEn_c$.
Compared with locally random dependency, the linear dependency map is more robust.
For the linear dependency map, the  percolation is always continuous(Fig.7).
Although the dependency distance changes strongly, the approximate entropies of
those dependency maps are almost equal to $0$. So their percolation
properties are nearly the same as that of a single lattice. It is thus clear that the randomness may be a more
important factor leading to cascade failures than dependency distance.
\begin{figure}
\includegraphics[width= 8.4cm]{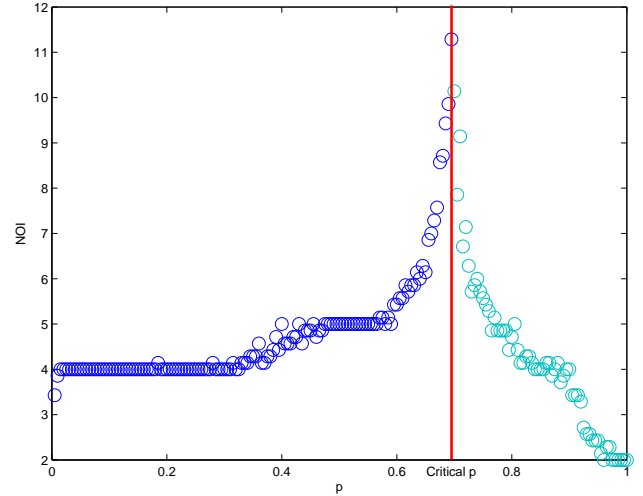}
\caption{The function of number of iterations(NOI) vs $p$ in interdependent lattice network($q = 1$).
The numerical results are obtained by averaging 100 realizations of networks consisting of $N = 10^4$ nodes.
The vertical red line is the critical line. On the left side of it, the system collapses down(blue circle), while
a giant component remains functional on the right side(green circle).
There is a sharp divergence of the NOI when $p$ approaches $p_c$. }
\label{fig5}
\end{figure}

\begin{figure}
\includegraphics[width= 8.4cm]{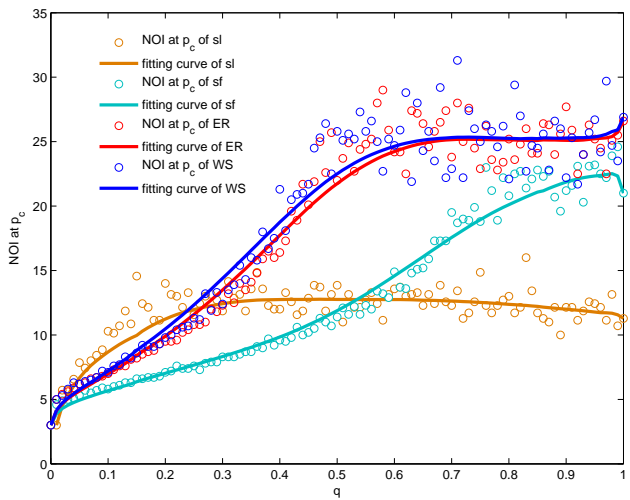}
\caption{The change of NOI at $p_c$ with $q$.
The brown, cyan, red and blue circles stand for NOI at $p_c$ of interdependent square lattice, scale-free
network, Watt-Strogatz network and Erd\H{o}s-R\'{e}nyi network respectively.
For interdependent random networks and small-world networks, NOI increases monotonously with $q$.
However for interdependent lattice network and scale-free network, there is one critical ${q_c}^{'}$.
When $q < {q_c}^{'}$, the NOI increases approximatively linearly, and when $q \geq {q_c}^{'}$, the $NOI$ starts to decreases.}
\label{fig6}
\end{figure}

Furthermore, it is possible that the randomness of dependency is related to other metrics of interdependent networks such as dimension.
The dimension of networks is a function of the distribution of link lengths\cite{daqing2011dimension}.
For spatially embedded networks, the dimension is one of the most fundamental quantities
to characterize its structure and very likely will influence its percolation property\cite{daqing2011dimension}.
However, how interdependency relationships between networks change the dimension of the system has not been figured out.
In reference \cite{li2012cascading}, the authors discover that the dependency length play an critical role in the
percolation transition. However, we found that under linear dependency map, the change of dependency length influence the percolation
property little. From the discrepancy of those two situations, it can be inferred that local property of dependency relationship
make a big difference. And the local property of dependency will directly influence the local topological inter-similarity between networks.
Randomness happen to reflects the local property of dependency (we can see this from the computation steps
of approximate entropy in Methods). In spatially interdependent embedded networks, the local characteristics of dependency
can be more intuitively characterized as the relative length of dependency links.
Under linear dependency map, the relative length of dependency links and the approximate entropy of dependency map  are nearly $0$.
No matter how large $r$ is, they changes little and are still nearly $0$, so the percolation changes little.
On the other hand, the smaller the relative length is , the less dimension is changed from that of single network.
There should exit some relations between dimension and the randomness of the dependency map.

\begin{figure}
\includegraphics[width= 8.4cm]{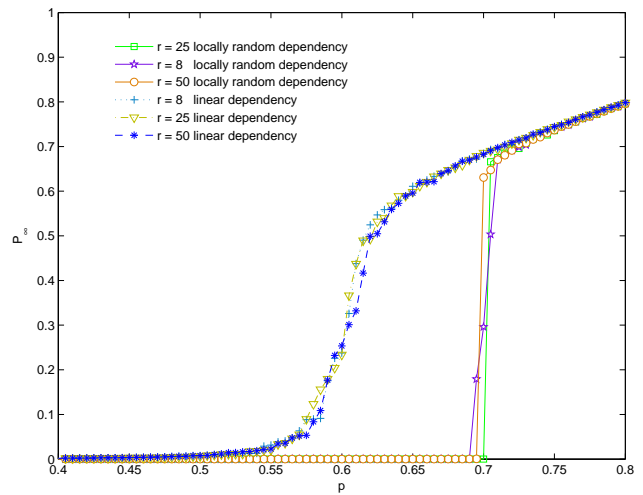}
\caption{The fraction of nodes in the giant component as a function of $q$.
For locally random interdependent network, when $r = 8$, the system represents the
characteristic of a second-order transition. For $r = 25$ and $r = 50$, the giant
component may completely collapse by removal of even a single additional node, which represents the characteristic of
a first-order transition. However for linearly interdependent networks, the transitions are second-order one when $r = 8,25,50$ and even $r = L$.}
\label{fig7}
\end{figure}

\section{Discussion}
In many real interdependent systems such as coupled power grid and communication
network, the dependency relationship is neither completely regular nor completely
random, but lies somewhere between these two extremes. The transition from regular
to random dependency is one of the keys to extreme vulnerability of spatially
interdependent systems. From the proposed intermediate cascade failure model (from regular to random dependency),
we find there is a transform from continuous percolation to discontinuous
percolation with the randomness variation of the dependency map between two
interdependent networks. We emphasize the generic character of our model because the
dependency map could influence not only the resilience but also synchronization, disease
spreading and other dynamic process in interdependent networks. With suitable
modification, our model could be applied to understand the dynamical process in
most real interdependent systems since the dependency maps between networks are more
various and complicated instead of totally random dependency or regular dependency.

The time scale of cascade failures is evidently important for system's resilience, but
it has received little attentions in the analysis of resilience so far. In different
dynamic processes, the characteristic time scales of systems vary greatly. For instance
biological systems, social systems and financial market dynamics have time scale much
longer than that of cascade failures of power grid.  Our analytic method based on
critical p is simple and effective for characterizing the time scale of different systems.
Generally, the system which has a shorter time scale demands higher requirements for our responding
speed to catastrophe and brings much bigger challenges for us to take mitigation actions than
those with longer time scale. Therefore, our method may provide a clue for research on
revealing the transient mechanism and mitigation of cascade failures in interdependent networks.

\section{Methods}

\subsection{Approximate entropy}
The randomness of the dependency maps the interdependent square lattice is measured by approximate
entropy. Entropy can effectively reflects the randomness of a sequence. However, for computation convenience, we choose
the approximate entropy as the measure of the randomness of the system. The approximate
entropy is denoted by $ApEn$ and is computed by following steps\cite{pincus1994physiological,pincus1991approximate}:

A. Given a series

$X(i) = [u(i), u(i+1),\cdots ,u(i+m-1)], i = 1 \sim  N - m + 1$

B. Count the distance between the vector $X(i)$ and other $X(j)$ for each i:

$d[X(i), X(j)] = \max \limits _{k = 0 \sim m-1}  |u(i+k) - u(j+k)|$

C. Given an threshold, count the the ratio between the number such that $d[X(i), X(j)] < r$ for each i
and the number of the vectors ie.$N-m+1$(devoted by $_i^m(r)$. as:

$C_r^m(r) =\frac{\{ the~ number~ of (d[X(i),X(j)]) < r\}}{N-m+1}$

Generally, $C_r^m(r)$ reflects m-dimensional pattern

D. $\phi^m(r) = (\frac{1}{N-m+1}) \sum \limits _{i=1}^{N-m+1}lnC_i^m(r)$

E. $ApEn(m,r) = \Phi^m(r) - \Phi^{m+1}(r)$

Parameter selection: here we  choose $m=2$ , and $r = 0.2* (standard~deviation~ of ~u)$.
\subsection{Percolation transition}
The percolation transition is studied by randomly removing a fraction $1-p$ of nodes and  the links
attached to them from both networks simultaneously. Then, on each network, clusters which are
detached from the largest connected component are removed. After that, the nodes in each network which
lost their supporting nodes in the opposite network are removed. This in turn causes more clusters to
break off from the giant component and this process is continued until no more clusters break off.
First, we analyze the situation with totally random dependency maps.
After the initial attack, only $p_1 = p_{\infty}(p)$ remains functional. Each node in A that is
removed causes the removal of its interdependent node in B. Then only $p_{\infty}(p_1)$
nodes in B remain alive. This produces further  damage in A
and  causes cascading failures. The cascade failures can be represented by the recursive
equation for the survived fraction $p_i$\cite{buldyrev2010catastrophic,li2012cascading}.
\begin{equation}
\begin{aligned}
    & p_0 = p,\\
    & p_1 = \frac{p}{P_0}p_{\infty}(p_0) = P_{\infty}(p),\\
    & \cdots\\
    & p_i = \frac{p}{P_{i-1}} p_{\infty}(p_{{i-1}}).
\end{aligned}
\end{equation}
In the limit $i \rightarrow \infty$, Eq.(1) yields the equation for the mutual giant component
at the steady state,
\begin{equation}
    x = \sqrt{pP_\infty (x)}.
\end{equation}
Equation (2) can be solved graphically by the intersection between the
curve $y = pP_\infty(x)$ and the straight line $y = x$.
Next, we consider the mutual percolation for more casual situations where the
dependency is not totally random. For every dependency link, there is a probability $q$
to rewire it at random. This is equivalent to the situation
with a fraction $1-q$ nodes mapping to itself and the remaining $q$ nodes
have a random dependency map. The case of  $q = 1$ corresponds to the scenario of a random dependency
map and $q = 0$ is identical to the conventional percolation on a single lattice.
For the initial attack which destroys a fraction $1-p$ of nodes, $\lfloor m = (1-p)N\rfloor$ nodes are removed.
We compute the probability $P_{same}$ that one node in   A depends on the same node in  B.
For $n$ nodes in totally random dependency networks, the number of nodes $E(n)$ in the same location
of both networks  is \cite{dorrie1958triumph}:
\begin{equation}
    E(n) = \frac{\sum \limits_{m =0}^n m * C_n^m D(n-m)}{n!},
\end{equation}
\begin{equation}
    D(n) = n!\sum \limits _{k =2 }^n {(-1)}^k * n!/k!.
\end{equation}
When $n$ is very large, the computation of  $D(n)$ is very inconvenient. For computation simplication, when $n \geq 2$, we have:
\begin{equation}
    D(n) \approx \lfloor n!/e + 0.5\rfloor.
\end{equation} where $e$ is the Euler's number and $\lfloor x \rfloor$ is the integer part of $x$. Then,
\begin{equation}
\begin{split}
    E(n) &  = \frac{\sum \limits _{m =1} ^ n \frac{m* n!}{(n-m)!m!} \lfloor (n-m)!/e+0.5)\rfloor}{n!}, \\
      & \approx \frac{\sum \limits _{m =1} ^ n \frac{m* n!}{(n-m)!m!}  (n-m)!/e+0.5)}{n!}, \\
      & = \sum _{m =1}^n \frac{1}{e(m-1)!} + \frac{1}{2(n-m)!(m-1)!},\\
      & = 1 + \sum \limits _{m = 0}^n \frac{1}{2(n-m+1)!m!} \leq 1 + \frac{e}{2}.
\end{split}
\end{equation}
So for each node, the probability that it is in the same location of A and B is:
\begin{equation}
    P_{same} = (1-q)*p + p*\frac{E(q*N)}{q*N}.
\end{equation}
When $n \rightarrow \infty$ , $P_{same} \rightarrow (1-q)*p$. The initial attack causes some number of nodes
to be disconnected from the giant component in both networks A and B. Furthermore, because of the dependency relationship,
the nodes disconnected from  A will lead to further damages.  $P_\infty$ increases with  $P_{same}$.
The greater $P_{same}$ is,  the more nodes disconnected from the giant component of A overlap the nodes in B. So
further damage decreases and cascade failures are weakened (or prevented) from the beginning . For $q = 0$, the
cascade failures are prevented from the start and the percolation is continuous.
When $q =1 $, the totally random dependency map will lead to a first-order transition.
As $q$ increases, the percolation transition changes from a continuous transition to a discontinuous one.
There must be one critical $q_c$ beyond which the percolation transition become discontinuous.

\begin{acknowledgments}
This paper is supported by the National Natural Science Foundation of China (Grant
Nos. 61170269, 61472045), the Beijing Higher Education Young Elite Teacher Project (Grant No.YETP0449),
the Asia Foresight Program under NSFC Grant (Grant No. 61411146001) and the Beijing Natural Science
Foundation (Grant No. 4142016).
\end{acknowledgments}

\nocite{*}

\end{document}